\documentstyle[11pt,aaspp4,psfig]{article}

\def\degree{^{\circ}}
\newcommand{\eg}{{\it e.g.\ }}
\newcommand{\ie}{{\it  i.e.\ }}
\newcommand{\etal}{et~al.\ }

\begin{document}

\long\def\comment#1{}

\title{Long-Term Stability of Planets in Binary Systems}

\author{Matthew J. Holman}

\affil{Harvard-Smithsonian Center for Astrophysics, 60 Garden Street, Cambridge MA 02138 USA}

\and

\author{Paul A. Wiegert}
\affil{Dept. of Physics and Astronomy, York University, Toronto Ontario M3J 3P1 CANADA}

\begin{abstract}

A simple question of celestial mechanics is investigated: in what
regions of phase space near a binary system can planets persist for
long times?  The planets are taken to be test particles moving in the
field of an eccentric binary system. A range of values of the binary
eccentricity and mass ratio is studied, and both the case of planets
orbiting close to one of the stars, and that of planets outside the
binary orbiting the system's center of mass, are examined.  From the
results, empirical expressions are developed for both $i)$ the largest
orbit around each of the stars, and $ii)$ the smallest orbit around
the binary system as a whole, in which test particles survive the
length of the integration ($10^4$ binary periods).  The empirical
expressions developed, which are roughly linear in both the mass ratio
$\mu$ and the binary eccentricity $e$, are determined for the range
$0.0 \le e \le 0.7$-$0.8$ and $0.1 \le \mu \le 0.9$ in both regions,
and can be used to guide searches for planets in binary systems.
After considering the case of a single low-mass planet in binary
systems, the stability of a mutually-interacting system of planets
orbiting one star of a binary system is examined, though in less
detail.

\end{abstract}


\newpage
\section{Introduction}

The long-standing question of whether planets can form and then
persist in binary star systems appears to have been answered
observationally.  Planets  have been detected about 55 $\rho^1$ Cancri,
$\tau$ Bootis and 16 Cygni B, all of which have companion stars
(Butler~\etal 1997; Cochran~\etal 1997).  Given that this basic
question has been answered, one might now wonder what types of binary
systems can harbor planets.

Rather than addressing the more difficult issue of how planets are
formed in binary systems, a simple question of celestial mechanics is
asked: in what regions of phase space near a binary system could
low-mass planets persist for long times?

An extensive body of literature is dedicated to this question; just a
few recent papers are noted here.  The papers by Szebehely (1980) and
Szebehely and McKenzie (1981) are examples of analytic work done on
this problem.  Graziani and Black (1981), Black (1982), and Pendleton
and Black (1983) used numerical experiments of three body systems to
develop empirical stability criteria.  H\'enon and Guyot (1970)
numerically studied the stability of periodic orbits in the circular
restricted problem as a function of the mass ratio.  Although these
studies were for generic systems, they were restricted to the case of
the stars following circular orbits.  Benest (1988; 1989; 1993; 1996)
explicitly included the eccentricity of the binary in his studies of
stability in the $\alpha$~Centauri, Sirius and $\eta$ Coronae Borealis
systems.  Dvorak (1984; 1986), Rabl and Dvorak (1988), and
Dvorak~\etal (1989) have studied the stability of test particles in
binary systems with equal mass stars as a function of the binary
eccentricity.

This body of literature has three principal limitations.  First,
although binaries typically have eccentric orbits, most of the
analytic results have been developed for the case of binaries with
circular orbits.  Second, numerical investigations of stability have
been restricted to special cases such as binary systems with equal
mass stars or particular binary systems.  Third, the numerical
investigations been limited to fairly short integrations. However, the
rapid increase in processor speed and improved algorithms make it a
straightforward task to extend the numerical results.

Orbits in binary systems have been traditionally separated into three
categories.  Following the designations of Dvorak and Rabl (1986), the
first includes planetary or P-type orbits.  These are well outside the
binary, where the planet essentially orbits the center of mass of the
stars.  The second type refers to the satellite or S-type orbits.
These are orbits near one of the stars, with the second star
considered to be a perturber.  The third type refers to orbits near the
$L_4$ or $L_5$ triangular Lagrange points.  These orbits are not
normally of interest for binary systems as the mass ratio must be less
than $\mu =m_2/(m_1+m_2) \approx 0.04$ for motion about these points
to be linearly stable.

In this study, both P and S-type orbits are examined. However, the
results for the S-type orbits, those close to one of the stars, are
more directly applicable to current extrasolar planet searches, as
radial velocity variations are more easily detected in systems in
which the planets are close to one of the stars.  It is our hope that
the results of our investigation can be used as a guide in selecting a
sample of suitable candidates for a radial velocity survey of binary
systems.

This work is different from that which precedes it in two ways: (1) a
full range of mass ratios has been examined, as well as a full range
of eccentricities; (2) these integrations extend for much longer
times than have been examined in the past.

In addition to looking at the stability of single planets in binary
systems, the stability of systems of multiple planets in eccentric
binaries is also investigated, albeit briefly. Interactions between
planets may affect their stability through angular momentum transfer
or other mechanisms.  The results of this part of the investigation
help us to understand the limits of results obtained for single
planets.

\section{Method}

This study investigates orbital stability numerically within the
elliptic restricted three body problem.  That is, the planets are
modeled as test particles moving in the gravitational field of a pair
of stars on fixed eccentric orbits about each other.  As the test
particles do not interact with each other, this approach allows us to
simulate large numbers of single planet systems simultaneously.

An approach first used by Dvorak (1986) is adopted here. First, an
eccentricity, mass ratio, and initial orbital longitude for the stars
is chosen.  Second, a battery of test particles on orbits in or near the
binary is started. These particles are on initially circular,
prograde orbits in the plane of the binary. This choice is based on
the results from the $\alpha$~Centauri system (Wiegert and Holman
1997) in which the largest stable orbit near the stars was found to
have an inclination in the plane of the binary\footnote{More
precisely, the largest stable orbits were those that were retrograde
in the binary plane, but these were only slightly (10\%) larger than
those at zero inclination.}.  For a range of semimajor axes, test
particles are started at eight uniformly spaced orbital longitudes.

Given the initial conditions of the binary system and the test
particles, the system is numerically integrated.  During the course of
the integration close encounters between the test particles and stars
are checked for, close encounters defined arbitrarily to be a passage
within $0.25$ of the binary semimajor axis $a_b$ of the non-central
star.  Escape orbits are also monitored.  Any test particle that
encounters one of the stars or escapes is removed from the
integration.  At the end of the integrations, the semimajor axis at
which the test particles at all initial longitudes survived the full
integration time is determined.  We call this the critical semimajor
axis \footnote{Rabl and Dvorak (1988) refer to this as the Lower
Critical Orbit.}.  The critical semimajor axis determines the
stability limit on the time scale of the numerical integration. Note
that this approach assumes that the critical semimajor axis is
well-defined (\ie that the boundary between stable and unstable
regions is sufficiently sharp); however, this will be shown to be a
valid assumption in most cases.

Binaries with eccentricities in the range $0.0 \le e \le 0.7$-$0.8$
and mass ratio in the range $0.1\le \mu \le 0.9$ have been examined,
though only the range $0.1 \le \mu \le 0.5$ has been studied in the
outer region, since systems with mass ratios of $\mu$ and $1-\mu$ are
equivalent in this regime (with the exception of a $180\degree$ change
in the particles' longitudes).  By the mass ratio $\mu = m_2/(m_1+ m_2)$ is
meant, where $m_1$ is the mass of the star the test particle is
orbiting and $m_2$ is the mass of the perturbing star.  For equal mass
stars $\mu = 0.5$. Both the eccentricities and mass ratios have been
explored in increments of $0.1$ or less. The initial phase of the
binary orbit is also a free parameter.  The two extremes are examined:
the binary initially at periapse and initially at apoapse, and the
more conservative result is taken (\ie that for which the stable
region is smallest), though the results do not differ significantly on
this account.  The initial conditions of our primary suite of
simulations are summarized in Table~\ref{table1}.

\begin{table}[tb]
\caption{Initial conditions for the binaries and test particles. The
binary semimajor axis is $a_b$, its eccentricity is $e$, its mass ratio
$\mu = m_2/(m_1 + m_2)$. A test particle's initial semimajor axis,
eccentricity, inclination relative to the binary plane, longitude of
the ascending node, argument of perihelion and mean anomaly are designated by $a$, $e_p$, $i$, $\Omega$, $\omega$ and $M$ respectively.}
\small
\centering
\begin{tabular}{cc}\hline\hline
Inner region & Outer region \\ \hline
\multicolumn{2}{c}{Binaries} \\ \hline
\multicolumn{2}{c}{$a_b=1.0$}  \\
$0.1 \le \mu \le 0.9$ & $0.1 \le \mu \le 0.5$ \\
\multicolumn{2}{c}{$\Delta \mu = 0.1$}\\   
$0.0 \le e \le 0.8$ & $0.0 \le e \le 0.7$ \\
\multicolumn{2}{c}{$\Delta e = 0.1$}\\ 
\multicolumn{2}{c}{Binary phase: periapse or apoapse} \\ \hline
\multicolumn{2}{c}{Test particles} \\ \hline
$0.02 \le a \le 0.5 a_b$ & $ 1.0 \le a \le 5.0 a_b$ \\
$\Delta a = 0.0025$-$0.01 a_b$ & $\Delta a = 0.1 a_b$ \\
\multicolumn{2}{c}{$e_p = i = \Omega = \omega = 0.0$} \\
\multicolumn{2}{c}{$M =  0\degree$,$45\degree$, $90\degree$, $135\degree$, $180\degree$, $225\degree$,
$270\degree$, $315\degree$}\\
\hline\hline
\end{tabular}
\label{table1}
\end{table}

For the orbital integrations the symplectic mapping method of Wisdom
and Holman (1991) was used.  Although this technique was developed for
numerically integrating systems of planets with a single dominant
central mass, it can also be applied to binary systems for orbits
close to one of the stars, when the other star is a distant perturber.
For orbits in the intervening region, where the forces from the two
stars are comparable, two techniques are used.  First, the symplectic
mapping method is employed, but with a smaller time step.  Second, a
conventional Bulirsch-Stoer integrator is used.  The region for which
the symplectic mapping method is not well-suited is also the region in
which test particles quickly escape, thus it is not necessary to use
an efficient method in this region.

\section{Single planet results: inner region}

Table~\ref{table2} shows an example of the results of the numerical
integrations in the inner region.  It lists the times survived by
test particles as a function of initial orbital longitude and
semimajor axis.  This example is for $\mu=0.5$ (equal mass stars) and
$e=0.5$.  The times are given in units of 10 binary periods.  A ``+''
marks those test particles that survived the full $10^4$ binary period
integration.  For this eccentricity and mass ratio the critical
semimajor axis is $a_c = 0.12$, as all test particles at or interior
to this value survived the full integration.  Those with the longest
removal times lie along the boundary, with removal time decreasing
outward.

\begin{table}[tb]
\caption{Planet survival times in the inner region, with $\mu= 0.5$
and $e = 0.5$.  Columns 1-8 show survival times for planets in
positions 1-8 with the stars started at periapse.  Column 9 shows the
initial semimajor axis of the planets in units where the binary
semimajor axis is unity.  Columns 10-18 give the survival times with
the stars started at apoapse.  All times are given in units of 10
binary periods.  A ``+'' indicates the test particle survived $10^4$
binary periods.  Here the critical semimajor axis is 0.12, the
farthest semimajor axis at which all the planets survived the full
integration.  Several of the test particles on the stability boundary
were terminated at times longer than the 300 binary periods integrated
by Rabl~and~Dvorak~(1988), but the stability boundary was not shifted
significantly by extending the integration to $10^4$ binary periods.
}

\small
\centering
\begin{tabular}{llllllllcllllllll}\hline\hline
\multicolumn{8}{c}{Binary initially at periapse}&   &\multicolumn{8}{c}{Binary initially at apoapse}\\ \hline
P1~~&P2~~&P3~~&P4~~&P5~~&P6~~&P7~~&P8~~&$a$~~&A1~~&A2~~&A3~~&A4~~&A5~~&A6~~&A7~~&A8~~\\\hline
 	3 	& 1 	& 1 	& 1 	&3 	&2 	& 1 	& 1 &0.17        & 1     &1     & 1     & 1     & 1     & 1     & 1     & 1 \\
 	7 	& 1 	& 1 	& 1 	&2 	& 1 	&2 	& 1 &0.16        & 1     &1     &1     &1     &1     &6     &1     & 1 \\
 	13 	&2 	& 1 	&1 	&5 	&2 	& 1 	&2 &0.15        &3     &2     &2     &1     &8     &2     &2     &2 \\
 	57 	&7 	&3 	&12 	&39 	&7 	&1 	&6  &0.14        & +     &9     &7     &4     &7     &18    &7     &13\\
 	192 	&159 	&11 	&106 	& + 	&53 	&62 	&56 &0.13        &776   & +     & +     &119   &267   &142   &511   & +  \\
 	 + 	& + 	& + 	& + 	& + 	& + 	& + 	& + &0.12        & +     & +     & +     & +     & +     & +     & +     & + \\
 	 + 	& + 	& + 	& + 	& + 	& + 	& + 	& + &0.11        & +     & +     & +     & +     & +     & +     & +     & + \\
 	 + 	& + 	& + 	& + 	& + 	& + 	& + 	& + &0.10        & +     & +     & +     & +     & +     & +     & +     & + \\
 	 + 	& + 	& + 	& + 	& + 	& + 	& + 	& + &0.09        & +     & +     & +     & +     & +     & +     & +     & + \\
 	 + 	& + 	& + 	& + 	& + 	& + 	& + 	& + &0.08        & +     & +     & +     & +     & +     & +     & +     &
+ \\\hline\hline
\end{tabular}
\label{table2}
\end{table}

Rabl and Dvorak (1988) concluded that an integration time of 300
binary periods was adequate to determine the gross stability boundary.
Here it is seen that several of the test particles on the stability
boundary were terminated at times longer than the 300 binary periods,
but the stability boundary was not shifted significantly by extending
the integration to $10^4$ binary periods.  Ten thousand binary periods
is still short compared to the ages of binary systems; therefore it is
possible that longer-term instabilities will arise.

The most important number derived from Table~\ref{table2} is the
critical semimajor axis.  For each pair of values of $\mu$ and $e$
this number is measured. These are listed in Table~\ref{table3}.  From
these numbers an empirical expression that gives the critical
semimajor axis as a function of $\mu$ and $e$ can be derived.  A least
squares fit to the data yields:
\begin{eqnarray}
a_c &=& [(0.464 \pm 0.006) + (-0.380 \pm 0.010)\mu +
(-0.631 \pm 0.034) e  \label{expression}\\
& & + (0.586 \pm 0.061)\mu e
+ (0.150 \pm 0.041) e^2 + (-0.198 \pm 0.074)\mu e^2] a_b,\nonumber
\end{eqnarray}\nopagebreak
where $a_c$ is the critical semimajor axis, $a_b$ is the binary
semimajor axis, $e$ is the binary eccentricity, and $\mu$ is the mass
ratio.  Each coefficient is listed along with its formal uncertainty;
the expression is valid to within 4\% (typically) and 11\%
(worst-case) over the range of $0.1 \le \mu \le 0.9$ and $0.0 \le e
\le 0.8$.  The expression was chosen to provide a reasonable fit with
a minimum of terms. It is arbitrary that the function is linear in
$\mu$ but quadratic in $e$.

\begin{table}[tb]
\caption{The critical semimajor axis in units of the binary semimajor
 axis for each pair of values of the mass ratio and eccentricity in
 the inner region. These empirical results are from the $10^4$ binary
 period numerical integrations.}  \centering
\begin{tabular}{clllllllllll} \hline\hline
&\hfill$\mu$&0.10 &0.20 &0.30 &0.40 &0.50 &0.60 &0.70 &0.80 &0.90 \\\hline
$e$ & \multicolumn{10}{l}{}\\  
0.0 & &0.45 &0.38 &0.37 &0.30 &0.26 &0.23 &0.20 &0.16 &0.13 \\
0.1 & &0.37 &0.32 &0.30 &0.27 &0.24 &0.20 &0.18 &0.15 &0.11 \\
0.2 & &0.34 &0.27 &0.25 &0.23 &0.20 &0.18 &0.16 &0.13 &0.10 \\
0.3 & &0.28 &0.24 &0.21 &0.19 &0.18 &0.16 &0.14 &0.12 &0.09 \\
0.4 & &0.23 &0.20 &0.18 &0.16 &0.15 &0.13 &0.11 &0.10 &0.07 \\
0.5 & &0.18 &0.16 &0.14 &0.13 &0.12 &0.10 &0.09 &0.08 &0.06 \\
0.6 & &0.13 &0.12 &0.11 &0.10 &0.09 &0.08 &0.07 &0.06 &0.045 \\
0.7 & &0.09 &0.08 &0.07 &0.07 &0.06 &0.05 &0.05 &0.045 &0.035 \\
0.8 & &0.05 &0.05 &0.04 &0.04 &0.04 &0.035 &0.03 &0.025 &0.0225 \\
\hline\hline
\end{tabular}
\label{table3}
\end{table}

\subsection{Hill's regime} \label{pa:hill}

\begin{figure}
\centerline{\psfig{figure=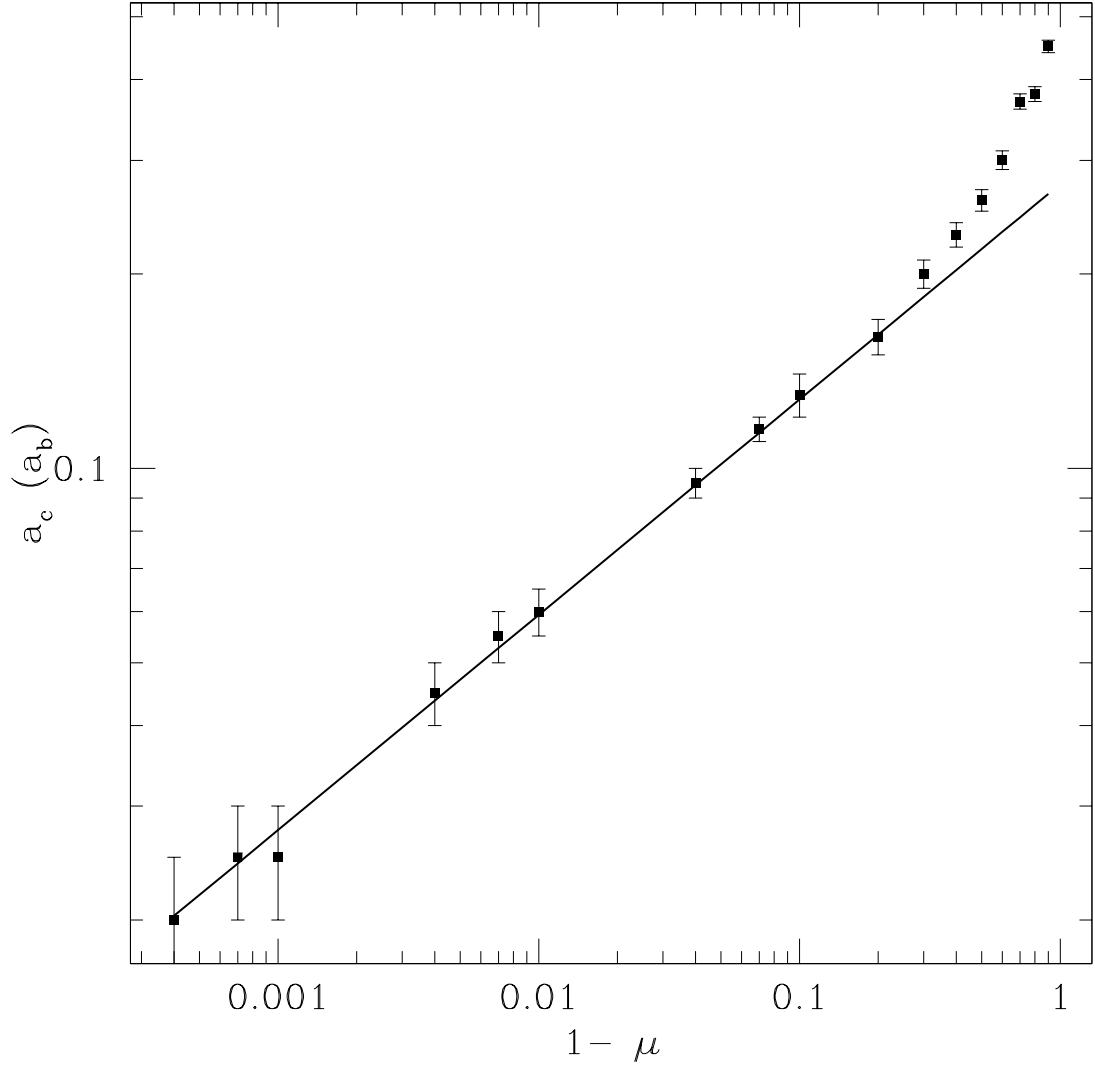,height=6in}}
\caption{The critical semimajor axis $a_c$ as a
function of $1-\mu$ for the binary eccentricity $e=0$. The heavy line
indicates the $(1-\mu)^{1/3}$ dependence of the Hill's sphere. The
error bars used in this and the following figures indicate the full
separation between simulated particles \ie if the particle separation
is $0.01 a_b$, the error bars span $\pm 0.01 a_b$.  \label{fi:hills}}
\end{figure}

As $(1-\mu)$ and $\mu \rightarrow 0$, one would expect the critical
semimajor axis to scale as with the Hill's sphere \ie as
$O(\mu^{1/3})$ or $O[(1-\mu)^{1/3}]$, an effect which is not predicted
by Eq.~\ref{expression}. In order to investigate, the simulations at
$e = 0.0$ were extended to a series of values in the range $0.9 \le
\mu < 1.0$. A plot of $a_c$ as a function of $\mu$ is shown in on a
log-log scale in Fig.~\ref{fi:hills}. The approximate linearity of
$a_c$ in $\mu$ changes to the expected $(1-\mu)^{1/3}$ dependence as
the mass of the primary and secondary diverge. These results imply two
distinct dynamical regimes for third-particle stability, one where
$(1-\mu)$ and $\mu \ll 1$ and one for $\mu \sim 0.5$, with the switch
occuring near where $(1-\mu)$ and $\mu \sim 0.2$.

\subsection{Comparison with earlier work}

\begin{figure}
\centerline{\psfig{figure=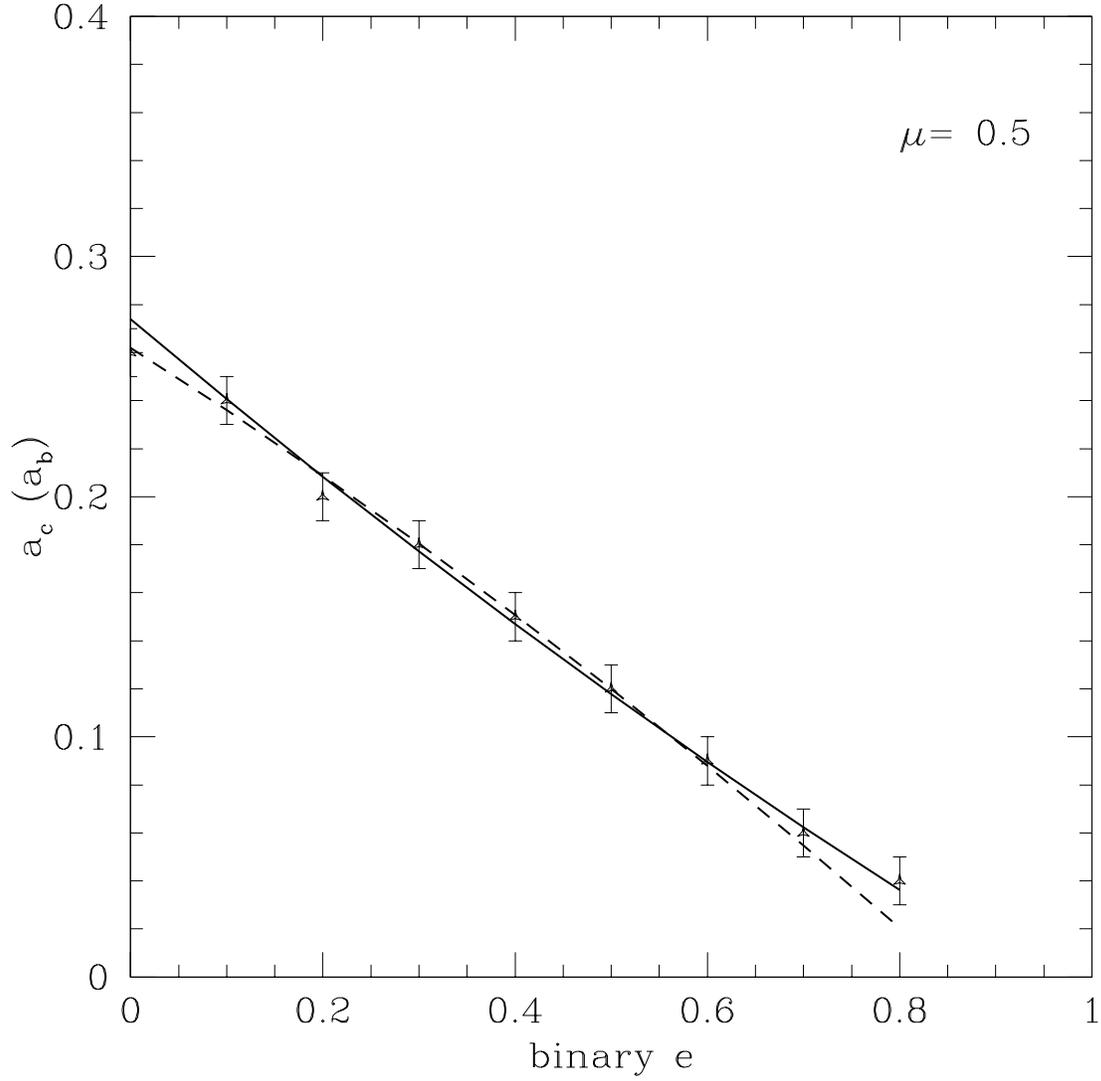,height=6in}}
\caption{The critical semimajor axis as a function
of the binary eccentricity for $\mu = 0.5$ (equal masses).  The solid
line is the least squares fit to results in this paper plotted for the
range of eccentricity studied.  The dashed line is the empirical fit
reported by Rabl and Dvorak (1988).  It is clear that the two are
consistent.  \label{fixed-e} }
\end{figure}

\begin{figure}
\centerline{\psfig{figure=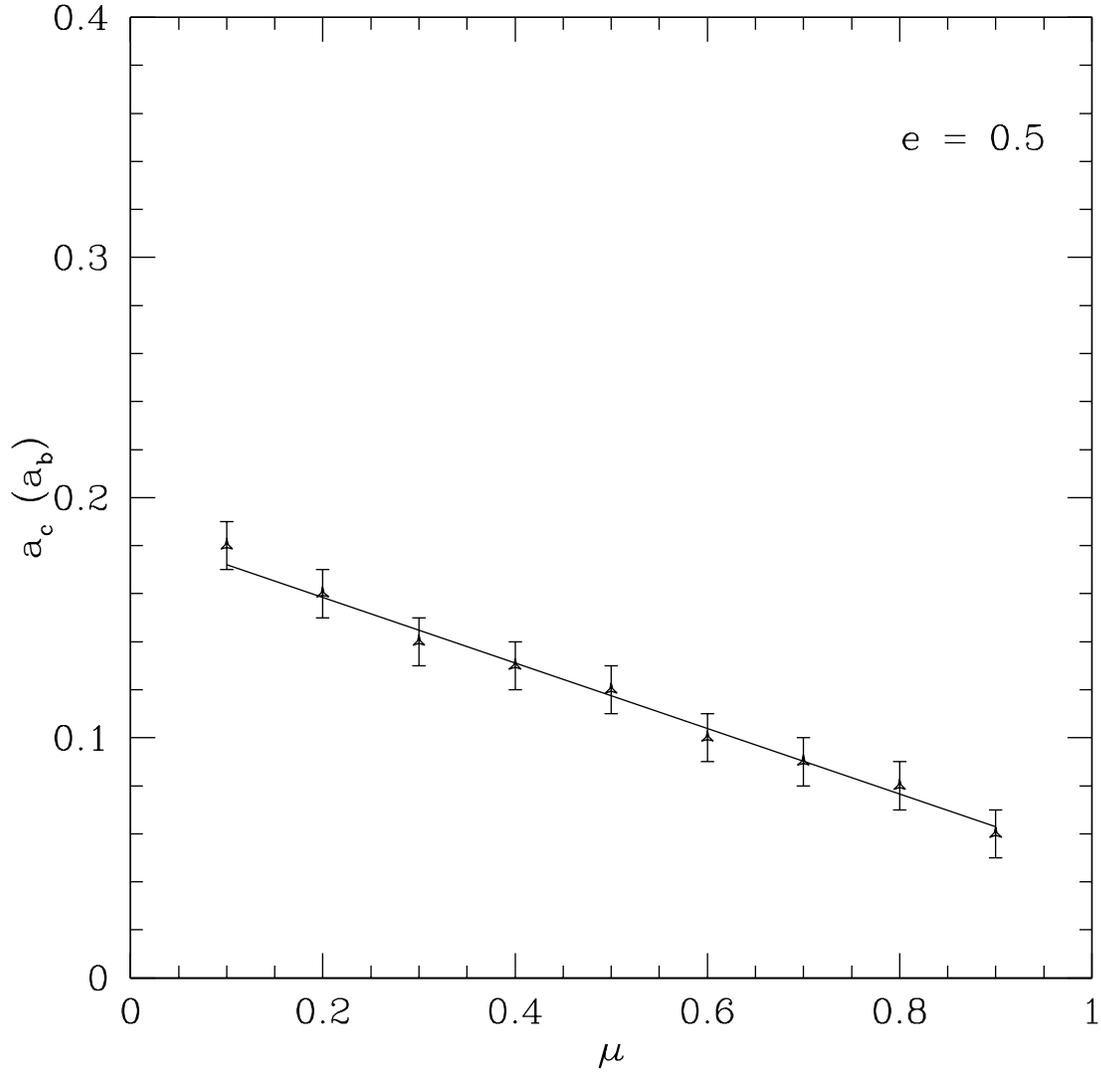,height=6in}}
\caption{The critical semimajor axis as a function
of the binary mass ratio $\mu$ for $e = 0.5$.  The least squares fit
is plotted for the range of mass ratios
studied. \label{fixed-mu} }
\end{figure}

Rabl and Dvorak (1988) report the following expression for the
critical semimajor axis in the case of equal mass stars:
\begin{eqnarray*}
a_c &=& [(0.262 \pm 0.006) + (-0.254 \pm 0.017) e + (-0.060 \pm 0.027)
e^2]a_b
\end{eqnarray*}
Inserting $\mu = 0.5$ into Eq.~\ref{expression} and collecting error
terms one gets:
\begin{eqnarray}
a_c &=& [(0.274 \pm 0.008) + (-0.338 \pm 0.045) e + (0.051 \pm 0.055)
e^2]a_b
\end{eqnarray}
In both cases it appears that a linear fit in $e$ would also work.
The two expressions are consistent, although our expression has a
somewhat sharper dependence on $e$.  Fig.~\ref{fixed-mu} shows the
data and least squares fit expression for the marginal case of
$\mu=0.5$.  The solid line is our least squares fit; the dashed line
is that of Rabl and Dvorak (1988).  They only explored up to $e=0.6$;
their line deviates most from ours when it is extrapolated beyond
$e=0.6$.  As longer integrations might tend to bring the entire curve
down, it is clear that this is not an important difference between the
investigations.  The fits are more strongly affected by the end points
than by the integration length. Fig.~\ref{fixed-e} shows the data and
empirical fit for the marginal case of $e=0.5$.  It is notable that
for this range of mass ratio the change in $a_c$ is very linear.  For
much larger or smaller values of $\mu$, the $O(\mu^{1/3})$ or
$O[(1-\mu)^{1/3}]$ scalings of the Hill's sphere occur
(\S~\ref{pa:hill}).

Aside from the work of Rabl and Dvorak (1988), there are other results
to which ours can be compared.  As noted above, Benest (1988, 1989,
1993, 1996) studied the $\alpha$~Centauri, Sirius and $\eta$ Coronae
Borealis systems (looking only at orbits in the plane of the binary).
We concentrate on his results for the $\alpha$~Centauri system. The
two stars in the tight binary of the $\alpha$ Centauri system have
masses $\mu_A = 0.54$ and $\mu_B=0.46$ and $e=0.52$.  Although Benest
uses a rotating-pulsating coordinate system that makes it difficult to
interpret his results, for $\alpha$~Cen~B he finds stability limits of
about $a_c = 0.19 a_b$ and $a_c = 0.15 a_b$ for the stars started at
periapse and apoapse, respectively. For $\alpha$~Cen~A, Benest only
gives the results for the stars started at periapse.  He finds $a_c =
0.23 a_b$.  Our empirical expression yields $a_c = 0.11 a_b$ and $a_c
= 0.12 a_b$, for $\alpha$~Cen~B and A, respectively.  It is unclear
why Benest's results indicate a much larger (up to a factor of two)
stable region than our own, when Rabl and Dvorak's, though only a
factor of three longer than Benest's, are so much closer to our
results. Perhaps the most unstable particles leave on time scales of
between one and a few hundred binary periods, with the remainder
evolving more slowly.

 Wiegert and Holman (1996) recently studied the $\alpha$ Centauri
system in detail, and examined the dependence of stability on
inclination.  For prograde orbits in the plane they find a limit of
$a_c = 0.11 a_b $, but used a much longer integration time ($3\times
10^5$ binary periods).  This suggests that the strongest instabilities
acting on time scales less than $\sim 10^6$ binary periods reveal
themselves rather quickly ($10^2$--$10^3$ periods); however, longer
time scales instabilities remain a possibility.

\subsection{Comparison with real binary systems}

In Table~\ref{table4} the binary systems listed in Dvorak~\etal (1989)
are re-examined.  For each system the period (years), parallax
(arcseconds), semimajor axis (AU), eccentricity, mass of the primary
(solar masses), the critical semimajor axis for test particles
orbiting about the larger mass, and the period (in years) of a test
particle at the critical semimajor axis.  Following that the mass of
the secondary star, its critical semimajor axis, and critical period
are listed, and then the values for the outer region
(\S~\ref{pa:outer}).  Our empirical expression (Eq.~\ref{expression})
has been used to determine the critical semimajor axes for these
systems.  The periods of the critical orbits follow simply from the
mass of the central star and the critical semimajor axis.

\begin{table}[tb]
\caption{Critical semimajor axes for the binary systems listed in
Dvorak {et al.} (1989).  For each system the period (in years),
parallax (in arcseconds), semimajor axis (in AU), and eccentricity are
listed.  Following the eccentricity we list the larger mass (in solar
masses) and the critical semimajor axis (in AU) and period (in years)
of a test particle orbiting about larger mass exactly at the critical
semimajor axis.  In the next six columns we list the mass, critical
semimajor axis, and critical period for the smaller mass in the
binary, and then the total mass along with $a_c$ and the period for
the outer (P-type) region. }

\centering
\begin{tabular}{lccccccccccccc}\hline\hline
Name &Period &Prlx.&    $a$&    $e$& $m_1$&      $a_{c1}$&$p_1$& $m_2$&       $a_{c2}$&$p_2$ &$M$ &$a_{cP}$& $p_P$ \\\hline
ADS 520         & 25.0&   0.07&  9.57&  0.22&  0.7&  1.93&  3.21&  0.7&  1.93&  3.21&1.40 & 29& 129\\
$\epsilon$ Cet  &  2.67&   0.069&  1.57&  0.27&  1.3&  0.29&  0.14&  1.3&  0.29&  0.14& 2.60 & 4.9& 6.7\\
$\gamma$ Vir    &171.37&   0.090& 37.84&  0.881&  0.94&  0.61&  0.49&  0.90&  0.60&  0.49& 1.84& 158& 1467\\
$\alpha$ Com    & 25.87&   0.038& 12.49&  0.5&  1.43&  1.49&  1.52&  1.37&  1.45&  1.49& 2.80& 45 & 182\\
$\epsilon$ CrB  & 41.56&   0.059& 13.98&  0.276&  0.79&  2.59&  4.68&  0.78&  2.57&  4.66& 1.57& 44 & 231\\
ADS 9716        & 55.88&   0.048& 19.15&  0.591&  1.135&  1.76&  2.20&  1.135&  1.76&  2.20& 2.27& 72 & 407\\
BD -$8^{\circ}$ 4352&  1.72&   0.152&  1.35&  0.05&  0.42&  0.35&  0.32&  0.42&  0.35&  0.32&0.84 & 3.4& 6.9\\
BD $45^{\circ}$ 2505& 12.98&   0.160&  4.58&  0.73&  0.285&  0.25&  0.23&  0.285&  0.25&  0.23&0.57 & 18& 103\\
ADS 11871       & 61.2&   0.061& 22.96&  0.249&  1.647&  4.49&  7.40&  1.582&  4.37&  7.26& 3.23& 71& 330\\
$\delta$ Equ    &  5.7&   0.052&  4.73&  0.42&  1.658&  0.67&  0.43&  1.593&  0.66&  0.42&3.25 & 16& 36.8\\
Kpr 37          & 21.85&   0.074&  9.67&  0.15&  1.2&  2.38&  3.35&  0.89&  1.96&  2.90&2.09 & 28& 102\\
99 Her          & 55.8&   0.060& 16.39&  0.74&  0.888&  0.97&  1.01&  0.522&  0.73&  0.87&1.41 & 68& 475\\
9 Pup           & 23.26&   0.067& 10.00&  0.69&  0.98&  0.67&  0.56&  0.87&  0.63&  0.54&1.85 & 40 & 184\\
ADS 15972       & 44.60&   0.248&  9.53&  0.41&  0.2728&  1.57&  3.76&  0.167&  1.17&  3.11& 0.44& 34& 302\\
$\alpha $ CMa   & 50.09&   0.375& 19.89&  0.592&  2.11&  2.17&  2.20&  1.04&  1.48&  1.76& 3.15& 79& 398\\
$\alpha $ Cen   & 79.92&   0.760& 23.57&  0.516&  1.12&  2.79&  4.40&  0.95&  2.54&  4.15& 2.07& 87& 567\\
$\xi $ Boo      &151.51&   0.147& 33.14&  0.512&  0.858&  3.96&  8.51&  0.731&  3.61&  8.02& 1.59& 123& 1076\\\hline\hline
\end{tabular}
\label{table4}
\end{table}

Although we make no statement about the process of forming planets in
binary systems, we note that the periods of small bodies orbiting at
the critical semimajor axis in these systems is longer than the
orbital periods of most of the recently discovered extra-solar
planets.

Returning to the recently discovered planets in binary systems, our
results can be compared with the sizes of the planetary orbits
found. The data summarised in Table~\ref{table5}. The size of the
binary orbits and eccentricities of these binaries are not
known. Their projected separation has been taken as the semimajor axis
and the critical semimajor axis has been computed for the extremes
$e=0$ and $e=0.8$. The orbits of the planets are within the critical
semimajor axis in all cases, however, the radial velocity measurements
used to find these planets are strongly biased towards planets very
close to their companion stars.
 
As stated, the results here concern the gross orbital stability of
planets orbiting in the plane of the binary system. Roughly speaking,
we have identified the regions of phase space in which the
perturbations in the with perturbations of the binary are strong
enough to significantly alter the semimajor axis of the
planet. However, the weak perturbations of a distant binary companion
can lead to large amplitude eccentricity oscillations for orbits that
are inclined more than about 40 degrees with respect to the binary
plane. Even when the orbital period of the binary is long enough
compared to that of the planet that the planets semimajor axis is
adiabatically preserved, the eccentricity oscillation can be large
enough to drive the planet at periapse into the surface of the central
star (Holman \etal 1997; Wiegert and Holman 1997).

\begin{table}[tb]
\caption{Basic data for the three binary systems thought to contain
planets, along with a comparison with the critical semimajor axes of
the systems with those of the planets (Hoffleit and Jaschek 1982;
Butler~et~al. 1997; Cochran~et~al. 1997). The binary's semimajor axis
is assumed to be equal to the projected radius. Some values are
estimates. Note that two planets have been detected in the 55
$\rho^1$~Cnc system.}

\centering
\begin{tabular}{lccc}\hline\hline
          & 16 Cyg B & 55 $\rho^1$ Cnc & $\tau$ Boo \\ \hline
HR number & 7504 & 3522 & 5185 \\
MK type   & G2.5V & G8V & F6IV \\
companion MK type & G1.5V & M2 & M2 \\
binary $\mu$ & $\sim 0.5$ & $\sim 0.8$ & $\sim 0.9$ \\
binary $r$ (AU, projected) & 1000 & 1200 & 100 \\
binary $e$ & ??? & ??? & ??? \\
planet $M \sin i$ ($M_J$) & 1.5 & 0.84, $>5$ & 3.87 \\
planet $a$ (AU) & 1.72 &0.11, $>4$ & 0.0462 \\
planet $\tau$ (days) & 800 & 14,$\sim 2800$ & 3.3 \\
planet $e$ & 0.63 & 0.05, ??? & 0.02 \\ \hline
$a_c$ (AU) if binary $e = 0.0$ & 260 & 200 & 13 \\
$a_c$ (AU) if binary $e = 0.8$ & 40 & 30 & 2.3 \\ \hline\hline
\end{tabular}
\label{table5}
\end{table}

\section{Single planet results: outer region} \label{pa:outer}

The results in the outer region are similar in some respects to those
in the inner region: the transition from the unstable region near the
binary to the stable region further outside occurs over a small range
in semimajor axis. For this reason, a simple empirical expression for
the stability boundary can be devised, as was done for the inner
region. However, an additional complication arises. In a large
fraction of the cases simulated, an additional island of instability
is seen beyond the inner ``boundary'' of the stable region. This
result is most easily seen by an example. Table~\ref{table6} shows the
results of a outer region simulation, in the format of
Table~\ref{table2} for $\mu=0.3$ and $e=0.4$. A island of instability
exists outside the inner unstable region: in this case, the island
arises near $a=3.7 a_b$. Such islands are seen in most of the
simulations.  They do not always occur at $a = 3.7 a_b$; rather, they
occur at the innermost $n:1$ mean motion resonance outside the inner
unstable region. Such instability islands are seen at the 3:1 through
the 9:1 mean-motion resonances under various circumstances (the
resonance revealed in Table~\ref{table6} is the 7:1), and have been
seen in previous studies (\eg H\'enon and Guyot 1970; Dvorak 1984;
Hagel and Dvorak 1988, Dvorak~\etal 1989).

\begin{table}[tb]
\caption{Planet survival times in the outer region, with $\mu= 0.3$, $e 
= 0.4$.  Columns 1-8 show survival times for planets in positions 1-8
with the stars started at periapse.  Column 9 shows the initial
semimajor axis of the planets in units where the binary semimajor axis
is unity.  Columns 10-18 give the survival times with the stars
started at apoapse.  All times are given in units of 1000 binary
periods.  A ``+'' indicates the test particle survived $10^4$ binary
periods.}

\small
\centering
\begin{tabular}{llllllllcllllllll}\hline\hline
\multicolumn{8}{c}{Binary initially at periapse}&   &\multicolumn{8}{c}{Binary initially at apoapse}\\ \hline
P1~~&P2~~&P3~~&P4~~&P5~~&P6~~&P7~~&P8~~&$a$~~&A1~~&A2~~&A3~~&A4~~&A5~~&A6~~&A7~~&A8~~\\\hline
0.25 & 0.25 & 0.43 & 0.81 & 0.21 & 1.1 & 0.75 & 0.75 & 3.1 &0.81 & 0.24 & 4.4& 2.7 & 0.48 & 0.4 & 2.3& 6.5\\
0.31 & 1.31 & 1.31 & 0.76 & 0.43 & 1.6 & 1.6 & 1.3 & 3.2 & 5.3 & 0.44 & 0.80 & 0.49 & 2.5 & 1.1 & 0.70 & 0.86\\
0.37 & 0.4 & 0.4 & 1.0 & 0.91 & 0.53& 3.8 & 0.56 & 3.3 &2.7 & 0.56 & 0.53 & 0.53 & 0.64 & 0.54 & 0.89 & 0.73 \\
2.5 & 0.76 & 0.67 &1.3 & 1.2 & 0.76 & 0.65 & 0.1.1 & 3.4 & + & 1.3 & 2.1 & 1.6 & + & 5.9 & 9.2 &0.67 \\
0.48 & 0.3 & 0.62 & 1.1 & 2.0 & 0.17 & 0.33 & 0.48 & 3.5 & + & + & + & + & + & + & + & + \\
 + & + & + & + & + & + & + & + & 3.6 & + & + & + & + & + & + & +  & + \\
2.9 & 1.8 & 1.8  & 2.3 & 0.45 & 1.2 & 1.1 & 6.2 & 3.7 & + & 4.1 & + & 6.4 & + & + & + & 1.35 \\
 + & + & + & + & + & + & + & + & 3.8 & + & + & + & + & + & + & + & + \\
 + & + & + & + & + & + & + & + & 3.9 & + & + & + & + & + & + & + & + \\
 + & + & + & + & + & + & + & + & 4.0 & + & + & + & + & + & + & + & + \\\hline\hline
\end{tabular}
\label{table6}
\end{table}

The phenomenon described above clearly implies that there is not a
sharp boundary between stable and unstable regions. It seem quite
possible that, if our simulations were extended, instability would
arise further and further from the central pair at higher and higher
order mean motion resonances. However, whether such resonances would
overlap sufficiently to render a large fraction of this outer
``stable'' region unstable is unclear.

Since the processors and numerical integration techniques at our
disposal do not allow us to resolve this situation through continued
simulations, we simply present the results obtained so far here, with
the caveat that further investigation is likely to prove them
incomplete. Since the region in which planets {\it could} survive is
of most interest here, the innermost semimajor axis at which planets
at all eight longitudes survive is chosen as our critical semimajor
axis. With this choice comes the clear possibility that some planets
with semimajor axes in excess of this value will be unstable. However,
since our goal is to provide a guide to observational searches and the
planets closest to the central stars will likely prove the easiest to
detect (by direct imaging surveys, say), one wishes to avoid placing
overly pessimistic limits on the stability region at this time. The
critical semimajor axes are listed in Table~\ref{table7}.

The critical semimajor axis is only a weak function of $\mu$, though
again roughly linear in $e$ over the range explored here. There is a
weak dependence on $\mu$ however, evidenced by a low broad peak at $\mu
\sim 0.25$. This peak is also seen in analyses of the Hill's stability
of the restricted three-body problem (Szebehely 1980; Szebehely and
McKenzie 1981). Because of this feature, up to second order terms in
$\mu$ are allowed in the fit in this case.

Our expression for the fit is
\begin{eqnarray}
\vspace*{-.5in}a_c & = & (1.60 \pm 0.04) + (5.10 \pm 0.05) e + (-2.22 \pm 0.11) e^2 + (4.12 \pm 0.09)\mu + \\
\vspace*{-.5in} && (-4.27 \pm 0.17) e \mu + (-5.09 \pm 0.11) \mu^2
 + (4.61\pm 0.36) e^2 \mu^2 \nonumber
\label{eq:outercrit}
\end{eqnarray}
The terms in $e^2 \mu$ and $e \mu^2$ have been omitted because their
addition provided little improvement of the fit. The residuals of
Eq.~\ref{eq:outercrit} show it to be good to 3\% typically, and to 6\%
worst-case over the range $0.0 \le e \le 0.7$ and $0.1 \le \mu \le
0.9$ in the outer region. Table~\ref{table4} lists the values of the
outer critical orbit for the real binary systems examined by
Dvorak~\etal (1989).

\begin{table}[tb]
\caption{The critical semimajor axis in units of the binary semimajor
 axis for each pair of values of the mass ratio and eccentricity in
 the outer region. These empirical results are from the $10^4$ binary
 period numerical integrations.}  \centering
\begin{tabular}{crlllll} \hline\hline
&$\mu$&0.10 &0.20 &0.30 &0.40 &0.50 \\\hline
$e$ && \multicolumn{5}{l}{}\\  
0.0 && 2.0 & 2.2 & 2.3 & 2.3 & 2.3 \\
0.1 && 2.4 & 2.7 & 2.7 & 2.8 & 2.8 \\
0.2 && 2.7 & 3.1 & 3.1 & 3.1 & 3.1 \\
0.3 && 3.1 & 3.5 & 3.5 & 3.3 & 3.2 \\
0.4 && 3.5 & 3.5 & 3.6 & 3.5 & 3.6 \\
0.5 && 3.8 & 3.9 & 3.9 & 3.6 & 3.7 \\
0.6 && 3.9 & 3.9 & 3.9 & 3.8 & 3.7 \\
0.7 && 4.2 & 4.3 & 4.3 & 4.1 & 4.1 \\
\hline\hline
\end{tabular}
\label{table7}
\end{table}

\subsection{Comparison with earlier work}

\begin{figure}
\centerline{\psfig{figure=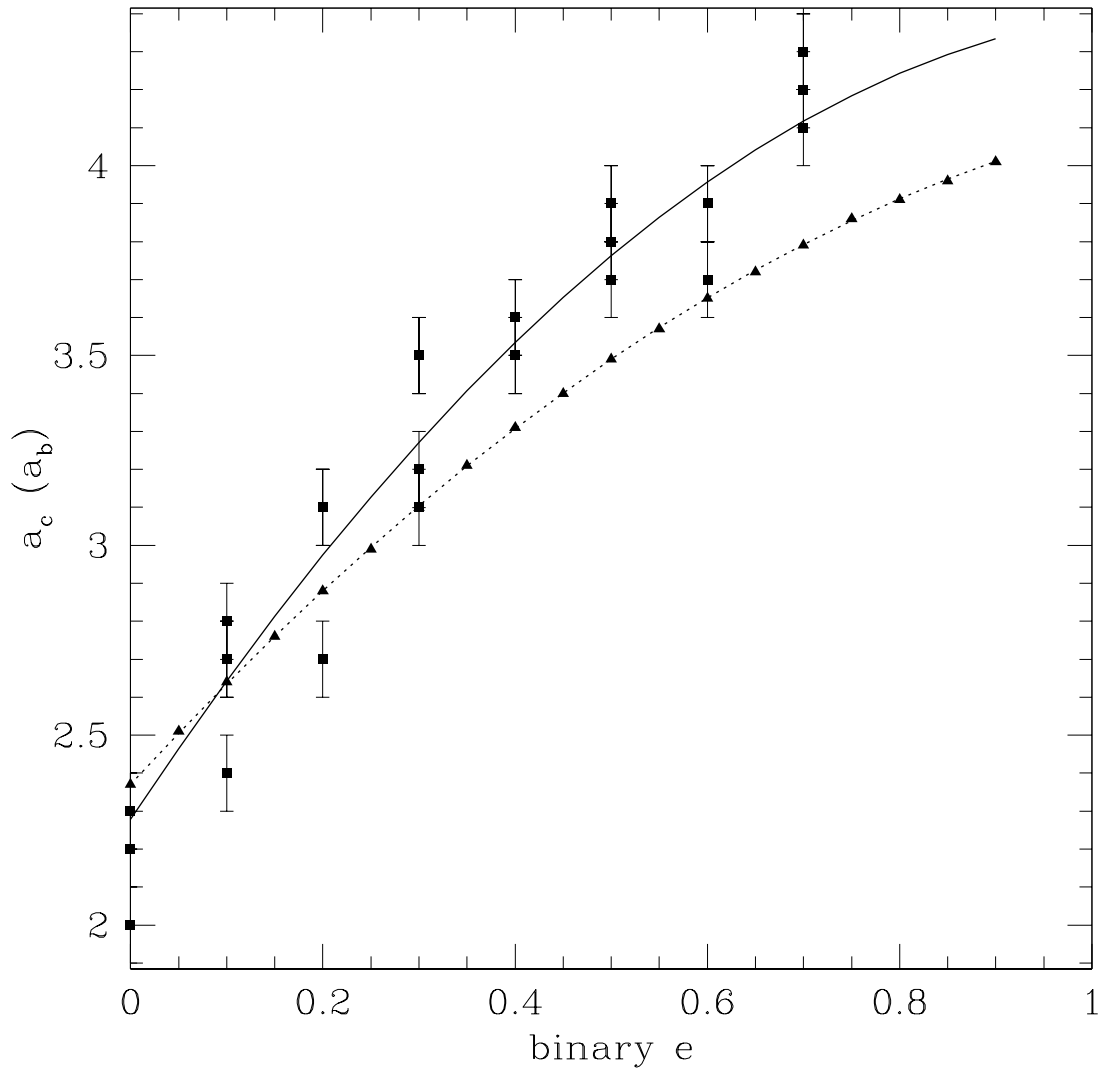,height=6in}}
\caption{The critical semimajor axis as a function
of eccentricity in the outer (P-type) region. The square data points
are our results (note: four different values of $\mu$ from 0.1 to 0.5
are used), the triangles, those of Dvorak~\etal (1989). Our
least-squares fit (up to quadratic in $e$, Eq.~\ref{eq:outercrit_e})
is shown by the heavy-line, that of Dvorak~\etal (see text) by a
dotted line.  Our simulations were run 20 times longer ($10^4$ binary
periods) and indicate that there is an erosion of the stable outer
region over time, at least at higher eccentricities. \label{fi:ofit}}
\end{figure}

Our results are very similar to those obtained by Dvorak~\etal (1989)
even though our simulations are 20 times longer. They find the
critical semimajor axis to be essentially independent of $\mu$, and
provide an expression for it in their Eq. (2). However, a comparison
of their expressions with their data leads us to suspect that their
Eqs. (1) and (2) may have their labels switched. Their Eq. (1) matches
their data for their ``upper critical orbit'' (our ``critical
semimajor axis'') much better than Eq. (2). If the equations are
mislabelled, their empirical expression is 
\begin{equation}
a_c = 2.37 + 2.76 e - 1.04 e^2.
\end{equation}
Performing a least-squares fit in the eccentricity, our data points
produce
\begin{equation}
a_c = (2.278 \pm 0.008) + (3.824 \pm 0.33)e - (1.71 \pm 0.10) e^2.
\label{eq:outercrit_e}
\end{equation}
with residuals indicating that this expression is within 5\% of the
experimental value typically, 14\% worst case. Dvorak~\etal do not
provide information on their residuals, but compare their empirical
expression with numerical results for a few independent (\ie real)
binary systems. They show slightly smaller errors (typical: within
2\%; worst case: within 10\%), but their error analysis is limited to
only seven independent cases, and only examined what they call the
Lower Critical Orbit (\ie the {\it largest} semimajor axis at which
all test particles become {\it unstable}). A comparison of our and
their data points with the empirical expressions is provided by
Fig.~\ref{fi:ofit}. We note that our data contains five points for
each value of $e$, corresponding to $\mu \in \{ 0.1, 0.2, 0.3 ,0.4,
0.5 \}$.

Our data points follow the same general trend as those of Dvorak \etal
but typically at slightly larger values.  This result indicates that,
though the edge between stable and unstable regions may not evolve
quickly, the stable region does erode as the time scales considered
lengthen, at least at higher binary eccentricities.

We also note that our results are comparable to Szebehely and
McKenzie's (1981) analytical results. They computed the Hill's
stability criterion for the circular restricted three-body case, and
found critical radii $r \approx 2.24$, 2.4, and $2.17 a_b$ for $\mu =
0.1$, 0.24, and 0.5. Our numerical results for $e=0.0$ provide $a_c
\approx 2.0$, 2.25, and $2.3 a_b$ for the same values of $\mu$.  The
closeness of these results confirms the validity of these numerical
integrations, and hints that the Hill's stability criterion may
provide more insight into the question of the stability of planets in
binary star systems.

Lastly, Wiegert and Holman (1997) found that the outer critical
semimajor axis $a_c = 3.7 a_b$ for central binary of $\alpha$~Cen
($e=0.52, \mu = 0.45$). This value is very close to the values derived
here, $3.73 a_b$ from Eq.~\ref{eq:outercrit} and $3.8 a_b$ from
Eq.~\ref{eq:outercrit_e}, though their integrations ran over three
times longer (32000 binary periods) than those presented here.

\section{Multiple planet results}

Now that stability limits have been derived for test particles
orbiting near one of the stars in a binary system, we consider systems
with multiple planets. Innanen~\etal (1997) examined the specific
effects of a binary companion on the orbits of the planets in our
Solar System.  We ask a similar but slightly different question: if
our own Solar System had a solar mass companion in an eccentric orbit
how large would its semimajor axis need to be for the planets to
survive $O(10^9$~yr)?

This question is addressed with numerical experiments.  In the first
set of experiments the system of our own Sun and the four giant
planets, Jupiter through Neptune, is augmented with a solar mass
companion and then numerical integrated for times up to one billion
years.  The companion was initially placed in the invariable plane of
the solar system with $\Omega = \omega = M = 0\degree$.  The
eccentricity of the companion was 0.4.  Several initial semimajor axes
of the companion were tested, and for each value of the initial
semimajor axis the system was integrated until the planetary orbits
began to cross or became hyperbolic with respect to the Sun.
Semimajor axes from $150$~AU to $500$~AU in increments of
$50$~AU.  For all companion semimajor axes except $400$~and~$500$~AU
Uranus and Neptune cross orbits within $10^7$--$10^8$~years.  In the
$400$~and~$500$~AU the systems survive the full $10^9$~year
integrations. That the $400$~AU system survived but the $450$~AU
system did not may indicate that particular configurations may have
additional stability arising from factors other than the distance of
the perturber.

In the second set of experiments we test the effect of a highly
inclined solar mass companion on a system of multiple planets.  Again
we start with our own Sun and giant planets.  To this we add a solar
mass companion inclined at $87\degree$ and $75\degree$ to the
invariable plane of the Sun and planets. The companion is started with
$\Omega = \omega = M = 0\degree$ and $e = 0.4$, and $a = 500$, $750$,
and $1000$~AU.  The $500$ and $750$~AU runs are unstable in
$10^7$--$10^8$~years, but the $1000$~AU runs survive $10^9$~years.  In
these runs the planets maintain their semimajor axes, eccentricities,
and relative inclinations, but the plane of the planets regresses
about the normal to the binary orbit.  The $i=75\degree$ run has a 110
Myr regression period; the $i=87\degree$ run has a 400 Myr regression
period.

We can compare the results of these experiments to those predicted by
Eq.~\ref{expression}.  For $a_b = 400$~AU, $\mu = 0.5$, and $e=0.4$ we
predict $a_c = 59$~AU.  Neptune's semimajor axis is $a_N = 30$~AU.
Our empirical expression overestimates the critical semimajor axis by
a factor of 2.  This is partly because, with a system of interacting
planets, it is not necessary to completely strip one of the planets
from its star to disrupt the system.  It is only necessary to perturb
the planetary eccentricities enough that the planets strongly
interact. Nevertheless, even with this factor of 2, it is reassuring
that the simple results from the elliptic restricted three body
problem predict the correct scale of the stability limit in our Solar
System.

\section{Summary and Future Work}

We have numerically investigated the long-term stability of planets
near one of the stars in a binary system.  Unlike earlier studies, we
have examined a full range of mass ratios and eccentricities.  From
the results we have derived an empirical expression for the semimajor
axis inside of which test particles on initially circular orbits
survive the full integration.  This expression is given as a function
mass ratio and semimajor axis. Our results for the inner S-type region
are confirmed, albeit weakly, by observations of planets in extrasolar
binary systems. Where our results examine parameter values studied by
earlier investigators, our results are largely consistent, despite our
significantly longer integration times. This fact indicates that the
stable/unstable boundary remains consistent over a wide range of time
scales, though some evidence for the erosion of the stable regions
over time has been seen. A further important proof of concept is that
systems of multiple planets, as modeled by our own outer planets, can
remain stable for times of order a billion years in the inner S-type region
of an eccentric binary system.


There are several avenues of future work in this field: 
\begin{itemize}
\item Here we have developed a very crude understanding of the limits
of stability.  Particular isolated regions of phase space near a
binary system, at the location of particular resonances, will have
different stability properties. The instabilities associated with the
$n$:1 mean-motion resonances (\S~\ref{pa:outer}), of which only a
cursory examination has been made here, are clear evidence of this.

\item
Despite the coincidence of our numerically computed stability boundary
with that derived analytically for Hill's stability by Szebehely and
McKenzie (1981), as yet we have no analytic derivation of the stability
limit, nor do we have a precise understanding of the source of the
chaos observed near and beyond the limit.    A semi-analytic
stability criterion based on the boundary between chaotic and
quasiperiodic motion is presented in Mardling and Aarseth (1998).

\item
The effects of giving the planets initial non-zero eccentricities, has
as yet been studied in only a very few cases (Benest 1988; 1989; 1996). 

\end{itemize}

\section{Acknowledgements}
We thank S. Tremaine, T. Mazeh and K. Innanen for helpful discussions.
We also thank J. Wisdom for the use of his computers. This work was
performed at the Canadian Institute for Theoretical Astrophysics and
the University of Toronto, and has been supported in part by the
Natural Sciences and Engineering Research Council of Canada.

\section{References}

\begin{description}

\item{Benest, D., 1988, {\sl Astron. Astrophys.} {\bf 206}, 143.} 
\item{Benest, D., 1989, {\sl Astron. Astrophys.} {\bf 223}, 361.} 
\item{Benest, D., 1993, {\sl Cel. Mech. Dyn. Astron.} {\bf 56}, 45.}
\item{Benest, D., 1996, {\sl Astron. Astrophys.} {\bf 314}, 983.}
\item{Black, D., 1982, {\sl Astron. J.} {\bf 87}, 1333.}
\item{Butler, R. P., Marcy G., Williams, E., Hauser, H. and Shirts,
P., 1997, {\sl Astrophys. J. Lett.} {\bf 474}, L115.}
\item{Cochran, W., Hatzes, A., Butler, P. and Marcy, G., 1997, {\sl
Astrophys. J} {\bf 483}, 457.}
\item{Dvorak, R., 1984, {\sl Cel. Mech.}  {\bf 34}, 369.}
\item{Dvorak, R., 1986, {\sl Astron. Astrophys.}  {\bf 167}, 379.}
\item{Dvorak, R., Froeschle, Ch., Froeschle, Cl., 1989, {\sl
Astron. Astrophys.}  {\bf 226}, 335.}
\item{Graziani, F. and Black, D., 1981, {\sl Astrophys. J.} {\bf 251},
337.}
\item{Hagel, J. and Dvorak, R., 1988, {\sl Cel. Mech.} {\bf 42}, 355.}
\item{H\'enon, M., Guyot, M., 1970, in {\it Periodic Orbits, Stability,
and Resonances}, ed. G.E.O. Giacaglia (Dordrecht:Reidel), p. 349.}
\item{Hoffleit, D. and Jaschek, C., 1982, {\sl The Bright Star Catalogue}, $4^{th}$ edition, (New Haven:Yale University Observatory) }
\item{Holman, M., Touma, J. and Tremaine, S., 1997, {\sl Nature} {\bf 386}, 254.}
\item{Innanen, K., Zheng, J., Mikkola, S., and Valtonen, M., 1997, {\sl Astron. J} {\bf 113}, 1915.}
\item{Mardling, R.A. and Aarseth, S.J. 1998, NATO ASI series, in press.}
\item{Pendleton, Y. J. and Black, D.C., 1983, {\sl Astron. J.} {\bf
88}, 1415.}
\item{Rabl, G., Dvorak, R., 1988, {\sl Astron. Astrophys.}  {\bf 191},
385.}
\item{Szebehely, V., 1980, {\sl Cel. Mech.} {\bf 22}, 7.}
\item{Szebehely, V. and McKenzie, R., 1981, {\sl Cel. Mech.} {\bf 23},
3.}
\item{Wiegert, P. A. and Holman, M. J., 1997, {\sl Astron. J.} {\bf
113}, 1445.}
\item{Wisdom, J. and Holman, M., 1991, {\sl Astron. J.} {\bf 102},
1528.}

\end{description}

\newpage

\end{document}